\documentstyle[preprint,epsfig,aps]{revtex}

\title{Cosmology of a brane radiating gravitons into the extra dimension}
\author{David Langlois, Lorenzo Sorbo, Mar\'{\i}a Rodr\'{\i}guez-Mart\'{\i}nez}
\address{GReCO, Institut d'Astrophysique de Paris, \\
Centre National de la Recherche Scientifique,\\
 98bis Boulevard Arago, 75014 Paris, France}
\date{\today}

\def\beq{\begin{equation}}
\def\eeq{\end{equation}}
\def\C{{\cal C}}

\def\hH{{\hat H}}
\def\hrho{{\hat \rho}}
\def\t{{\hat t}}
\def\hC{{\hat\C}}
\def\F{{\sigma}}
\def\p{{p}}
\def\x{{\bf x}}
\def\M{{M}}

\begin{document}

\maketitle

\begin{abstract}
We study in a self-consistent way 
the impact of the emission of bulk gravitons on the 
(homogeneous) cosmology of a three-brane embedded in a five-dimensional 
spacetime. In the low energy regime, we recover the well known result 
that the bulk affects the Friedmann equation only via a radiation-like 
 term $\C/a^4$, called dark or Weyl radiation. By contrast, 
in the high energy regime, we find that the Weyl parameter $\C$ is no 
longer constant but instead grows very rapidly as $\C\propto a^4$. 
As a consequence, the value of $\C$ today is not a free parameter as 
usually considered  but is a fixed number, which, generically, depends
only on the number of relativistic degrees of freedom at the high/low 
energy transition. Our estimated amount of Weyl radiation satisfies the 
present nucleosynthesis bounds.
\end{abstract}

\section{Introduction}
In the last couple of years, considerable attention  has been paid to  
a new type of models with extra dimensions, for which 
ordinary matter is confined on a three-dimensional space called brane.
Although a lot of effort has been devoted to investigate the signatures 
of extra-dimensions in gravity tests and collider experiments, 
these  potentially new effects will be inaccessible if
 the higher-dimensional fundamental mass is too high.
Cosmology would then be the last hope to detect an extra-dimensional 
signature, via for instance the Cosmic Microwave Background 
Radiation (CMBR) anisotropies \cite{l00b}, 
 because of the high energies reached in the very early universe. 

The simplest viable model of brane cosmology that 
takes into account the {\it self-gravity} of the brane is 
the cosmological extension \cite{cosmors} 
of the Randall-Sundrum (RS) setup \cite{rs99b}
 and consists of a $Z_2$-symmetric 
(i.e. mirror symmetric) three-brane, with a tension $\lambda$ and 
containing ordinary cosmological matter with energy density $\rho$, 
 embedded in an {\it empty} five-dimensional spacetime endowed with a 
(negative) cosmological constant $\Lambda\equiv -6\mu^2$.
In this context, one can solve explicitly \cite{bdel99}, 
for {\it any equation of state} 
of brane matter,
the   five-dimensional  Einstein's equations
\beq
G_{AB}+\Lambda g_{AB}=\kappa^2 T_{AB},
\label{einstein}
\eeq 
where $\kappa^2\equiv \M^{-3}$ is the five-dimensional gravitational coupling
and 
where the energy-momentum tensor here vanishes everywhere except on the brane.
Imposing the RS condition \cite{rs99b} $\kappa^2\lambda=6\mu$,  this
 leads to the  generalized Friedmann law \cite{bdel99,kraus99,sms99}
\beq
H^2= {\kappa^4\over 18}\lambda\rho
+{\kappa^4\over 36}\rho^2+{\C\over a^4},
\label{fried}
\eeq
where $H\equiv \dot a/a$ is the expansion rate and 
 $\C$ is an arbitrary constant.
Although the first term on the right-hand side, with the identification 
$\kappa^4\lambda/6=\kappa^2\mu\equiv 8\pi G={\bar M}_p^{-2}$, 
agrees with 
the familiar Friedmann equation, the two additional terms induce a   
deviation from the usual law. 
First, there is a term quadratic in $\rho$ \cite{bdl99}, 
which dominates the usual 
linear term in the {\it high energy regime}, when $\rho\gg\lambda$.
Second, one finds a radiation-like term proportional to 
the  integration constant $\C$, which we will call here the 
{\it Weyl parameter} since it is related to the bulk
 Weyl tensor \cite{sms99} and which is formally analogous  
to the Schwarzschild mass which arises as an integration constant when solving 
the vacuum Einstein equations with spherical symmetry. Indeed,  
the five-dimensional spacetime metric is Schwarzschild-Anti de Sitter
(Sch-AdS) which 
can be written in the manifestly static form \cite{kraus99}
\beq
ds^2=-f(r)dT^2+{dr^2\over f(r)}+r^2 d\x^2,\quad 
f(r)=\mu^2 r^2-{\C\over r^2},
\label{ads}
\eeq 
and the `radial' coordinate $r$ of the brane can be identified with its scale
factor $a$.   
When the bulk spacetime is empty, $\C$ is necessarily constant in time. This 
can be seen as a generalization to five dimensions of Birkoff's theorem 
\cite{bcg00}. 

In this setup, one recovers the usual conservation law $\dot\rho
+3H(\rho+p)=0$. This picture is however oversimplified in the sense that, 
if the assumption of homogeneity is justified on average, there exist 
fluctuations on small scales, which create gravitational waves that can 
escape into the bulk. In the language of particle physics, one would 
say that interactions of brane particles will generate bulk gravitons. 
If this question  has been analysed in detail in the context of flat 
compact extra-dimensions (see e.g. \cite{add}), 
it has received so far  much less attention in the RS context 
\cite{hm01}, the main difficulty being to take into account the 
back reaction of the graviton flow on the bulk metric. 
The purpose of the present work is precisely to present 
a  treatment that  combines  {\it self-consistently} the emission 
of gravitons with  the surrounding  spacetime in which our brane-universe is 
moving.

\section{Brane in a radiative spacetime}
In order to  model the bulk spacetime metric, 
we  generalize to  five dimensions the four-dimensional   
Vaidya's metric \cite{vaidya}, 
which is used in  general relativity  to   describe the spacetime 
surrounding a radiating star. 
Our metric ansatz  is  given by
\beq
ds^2=-f\left(r,\,v\right)\,dv^2+2\,dr\,dv+r^2\,d\x^2, \quad 
f(r,v)=\mu^2 r^2-{\C(v)\over r^2}.
\label{metric}
\eeq
If $\C$ does not depend on $v$, the above metric is 
simply a rewriting of the Sch-AdS metric (\ref{ads}) 
with the change of coordinate 
$v=T+\int dr/f(r)$. In all cases, $v$ is a null coordinate since the 
spacetime trajectory defined by $v=const$ (and $x^i=const$) is  light-like.
In Schwarzschild spacetime, it would correspond to {\it ingoing} radial 
light rays. Note that, strictly speaking, the Vaidya spacetime corresponds
to an {\it outgoing} radiation flow, whereas in our case, we are interested 
in an {\it ingoing} radial flow because the brane, which emits the radiation, 
is in some sense located at the largest radius of spacetime (when 
getting away from the brane, one moves radially `inwards', as can be deduced 
from the junction conditions (\ref{israel})).
 
In the above spacetime (\ref{metric}), the trajectory of the brane can be  
expressed in terms of 
 its coordinates $v(t)$ and $r(t)$ as functions of the 
proper time $t$ (which is also the cosmic time). 
The normalization of the velocity vector ($u^A=\{\dot v, \dot r, {\bf 0}\}$, 
where a dot stands for 
a derivation with respect to $t$) implies
\beq
\dot v={\dot r+\sqrt{f+\dot r^2}\over f},
\label{vdot}
\eeq
 assuming  that the brane is always in the region of spacetime
where $f>0$ (i.e.  outside the  horizon in the Schwarzschild case).

Similarly to  its four-dimensional counterpart, it is easy to verify that the 
metric (\ref{metric}) is a solution of Einstein's equations (\ref{einstein})
with a  bulk energy-momentum tensor of  the form
\beq
T_{AB}=\sigma k_A k_B,
\label{Tbulk}
\eeq
where $k^A$ is a  (here ingoing) null vector. 
Our model thus corresponds to a simplified picture, where the bulk 
gravitons are supposed to escape `radially' from the brane.
 Choosing to normalize 
it  so that $k_A u^A=1$, the only non-vanishing component 
is 
$k^r (=k_v) =1/\dot v$. The quantity $\sigma$ represents, 
for a brane observer,  the energy flux of the bulk gravitons and 
is  related, via Einstein's equations (\ref{einstein}),  
to the variation  of the
Weyl parameter, according to the expression
\beq
{d\C\over dv}={2\kappa^2\sigma\over 3}r^3
\left(\dot r-\sqrt{f+\dot r^2}\right)^2.
\label{sigma}
\eeq

The  Einstein equations being solved in the bulk, it remains to  
impose at the brane location the  Israel junction conditions, which  
relate the jump of the 
 extrinsic curvature tensor for the brane $K_{AB}=h_A^C\nabla_C n_B$
($n^A$ is the unit vector normal to the 
brane and $h_{AB}=g_{AB}-n_An_B$ the induced metric) to the brane 
energy-momentum tensor $\tau_{AB}$, namely \cite{bdl99}
\beq
\left[K_{AB}\right]=\kappa^2\left(\tau_{AB}-{1\over 3}\tau h_{AB}\right).
\label{israel}
\eeq
Here,
$\tau_{AB}=(\rho_b+p_b)u_A u_B+p_b h_{AB}$, where
 $\rho_b=\lambda+\rho$ and $p_b=-\lambda+p$ are 
respectively the {\it total} energy density and pressure in the brane. 
The ordinary spatial components of the junction conditions yields 
(using  $n_A=\{\dot r, -\dot v, {\bf 0}\}$) the expression
\beq
{\kappa^2\over 6}\rho_b={\sqrt{f+{\dot r}^2}\over r},
\label{junction}
\eeq
which implies the same unconventional Friedmann equation as in (\ref{fried}),
 with the 
only difference that $\C$ is a function of $v$ instead of a constant.
The other  components (along the time and the fifth dimension) 
yield another relation, which 
upon using (\ref{fried}) and its time derivative as well as 
(\ref{sigma}), can be rewritten as  
\beq
\dot \rho_b+3{\dot r\over r} (\rho_b+p_b)=-2\sigma.
\eeq
This equation simply expresses the fact that, from the point of view 
of the brane, the energy loss suffered by the brane   
exactly corresponds  to the bulk radiation flow embodied in 
(\ref{Tbulk}) (note that the right-hand side is very simple here because
we have chosen to normalize $k^A$, and thus $\sigma$, with respect to the 
brane observers).
 There is a  factor $2$ because  the brane is radiating on both sides, 
into  {\it two} copies 
of the same bulk spacetime.

\section{Production rate of bulk gravitons}
In order to determine quantitatively the energy loss $\sigma$
 we now need to evaluate the cross-section 
of the  process $\psi+\bar\psi\rightarrow G$, where $\psi$ is a particle 
confined on the brane ($\bar\psi$ its antiparticle) 
and $G$ a bulk graviton, process which appears to  
be the dominant effect in the very early universe.
To compute this cross-section, one can ignore the cosmological influence 
since the temperature $T$ of the brane particles is always much bigger than 
$H$ (indeed $H\sim T^4/\M^3$ in the high energy regime and 
$H\sim T^2/{\bar M}_p$ 
in the low energy regime, and we always assume $T\ll \M$).
A  Minkowski background {\it in the brane}  corresponds  
 to the five-dimensional Randall-Sundrum  solution \cite{rs99b}. 
Starting from  the linear perturbations of the RS metric, 
\beq
ds^2=\left(e^{-2\mu  |y|} \,  \eta_{\mu\nu}+ 2\kappa 
h_{\mu\nu} \left(x,\,y\right) \right)dx^\mu\,dx^\nu+dy^2\,\,,
\eeq
in the RS gauge (i.e.  $h^\mu{}_\mu=0$ and 
$\partial^\mu h_{\mu\nu}=0$), the gravitons 
correspond to the decomposition  
into (generalized) Kaluza-Klein (KK) modes, 
\beq
h_{\mu\nu}\left(x,\,y\right)=\int dm\,  u_m(y) \phi^{(m)}_{\mu\nu}(x), 
\eeq
 where the modes $u_m(y)$ are given by  \cite{gt99}
\beq
u_m(y)=\sqrt{m\over 2\mu}{J_1(m/\mu)N_2\left(me^{\mu |y|}/\mu\right)- 
N_1(m/\mu)J_2\left(me^{\mu |y|}/\mu\right)
\over \sqrt{\left(J_1(m/\mu)\right)^2+\left(N_1(m/\mu)\right)^2}},
\label{um}
\eeq
and 
satisfy  the normalization $\int \, dy\, e^{2\mu|y|}\, u_m^*(y) u_{m'}(y)=\delta(m-m')$.
The $h_{\mu\nu}$ have been defined so that the kinetic part of the 
graviton action is of the canonical form $S_{kin}=-(1/2)\int dm\int d^4x 
(\partial\phi_m)^2$. 
The interaction between brane matter (living at $y=0$) and the 
bulk gravitons is described by the action 
\beq
{\cal S}_{int}
=\kappa\int dm \, 
u_m(0)\int d^4 x \, \tau^{\mu\nu}\, \phi^{(m)}_{\mu\nu},
\label{int}
\eeq
which means that the effective coupling constant for the `canonical' KK modes
$\phi_m$ is $\kappa |u_m(0)|$.
For the modes $m\gg \mu$,  (\ref{um}) gives us  
$|u_m(0)|\simeq 1/\sqrt{\pi}$. 

The interaction  Lagrangian (\ref{int}) 
gives us the explicit coupling between any bulk 
KK graviton and brane matter, from which one can derive the amplitude 
for the scattering of brane particles leading to a KK emission.
At this stage, the calculation is quite analogous to the procedure 
already described in the context of flat extra dimensions \cite{feynman}, 
the only difference being the coupling constant in (\ref{int}).
Using those results, one finds that the   
spin and particle-antiparticle averaged squared amplitude 
(neglecting the  mass of the  incoming particles) is given here by 
\begin{equation}\label{ampli}
\sum \left\vert {\cal {M}}\right\vert^2 = \kappa^2 |u_m(0)|^2 A\, {s^2\over 8}, 
\end{equation}
with  $s=\left(\p_1 + \p_2 \right)^2$ 
($\p_1$ and $\p_2$ being the incoming 4-momenta of the scattering particles), 
and with $A_s=2/3$ for scalar-scalar scattering, $A_f=1$ for 
fermion-fermion scattering, and $A_v=4$ for photon-photon scattering.

Going back to cosmology, the brane matter  energy density  deficit 
corresponds to the energy density created in the form of bulk 
gravitons, which can be expressed, after integrating  the  Boltzmann 
equation over  momentum space, as 
\beq
{d\rho \over dt}+ 3H(\rho+p)=-\int dm \int\frac{d^3p_m}{\left(2 \pi\right)^3}
\, {\bf C}_m,
\label{boltzmann}
\eeq
where the collision term associated with the scattering process 
described above  is given by
\begin{equation}
{\bf C}\left[f\right]={1\over 2}
\int \frac{d^3p_1}{\left(2 \pi\right)^3\,2E_1}\, \frac{d^3p_2}{\left(2 \pi\right)^3\,2E_2}\,\sum \left\vert {\cal {M}}\right\vert^2\,f_1\,f_2\,\left(2\pi\right)^4\,\delta^{(4)}\left(\p_1+\p_2-\p_m\right)\,\, ,
\end{equation}
with the Fermi/Bose  distribution functions $f_i=1/(e^{E_i/T}\pm 1)$, while  
 $\p_m$ is the four-momentum of the created bulk graviton. 
We  will assume that the temperature  
satisfies  $T\gg\mu$ (this is an excellent approximation in the early 
universe for a large range of values for $\mu$ 
since the only constraint, from 
gravity experiments, is $\mu\gtrsim  10^{-3}$ eV ). 
The typical energy of gravitons created by brane particles will 
be of the order of $T$, 
so that most of the contribution to the energy loss will 
come from `heavy' gravitons satisfying $m\gg \mu$. Therefore, 
we  simplify the calculation by  directly replacing 
$u_m(0)$ by $1/\sqrt{\pi}$ in the right-hand side of (\ref{boltzmann}), thus
 introducing only a slight 
error due to the very light 
KK modes ($m\lesssim \mu$).  
One can then perform explicitly the integration to
find
\beq
\dot\rho+4H\rho=-\left[{315\zeta(9/2)\zeta(7/2)\over 512\pi^3}\right]
\, \hat g(T) \, \kappa^2\,  T^8,
\eeq
where we have introduced an {\it effective} number of 
degrees of freedom (relevant for bulk graviton production) defined by 
\beq
\hat g(T)=\left((2/3)g_s+4g_v +\left(1-2^{-7/2}\right)\left(1-2^{-5/2}\right)
g_f\right),
\eeq
where $g_s$, $g_f$ and $g_v$ are the 
number of relativistic degrees of freedom, at temperature $T$,  for particles 
of  respectively  
spin $0$, $1/2$ and $1$.  

\section{Coupled cosmological evolution}
 We now combine  all the above results 
 in order to get a dynamical system describing the {\it coupled 
evolution} of the brane energy density, of the brane scale factor and 
of the Weyl parameter.

In the radiation era, on which we focus, the energy density loss rate 
$\F$ can be rewritten as 
\beq
\F={\alpha\over 12}\kappa^2\rho^2, \quad \alpha=
{ 212625\over 64\pi^7}\zeta(9/2)\zeta(7/2){\hat g\over g_*^2},
\eeq
after  use of  the relation $\rho=(\pi^2/30)g_*T^4$, where 
$g_*=g_s+g_v+(7/8)g_f$ is the total effective number of relativistic 
degrees of freedom in the energy density. If all degrees of freedom of the 
standard model are relativistic, $g_*=106.75$ and $\hat g\simeq 166.21$ 
so that 
$\alpha\simeq 0.019$.
Introducing the dimensionless quantities  $\hrho=\rho/\lambda$, 
$\hH=H/\mu$, $\t=\mu t$, the equation governing the evolution of the 
energy density now reads
\beq
{d\hrho\over d\t}+4\hH\hrho=-\alpha\hrho^2,
\label{rhodot}
\eeq
whereas the Friedmann equation takes the form 
\beq
\hH^2=2\hrho+\hrho^2+{\hC\over a^4},
\label{adot}
\eeq
with  the dimensionless Weyl parameter $\hC=\C/\mu^2$.
The final  equation we need 
is that governing the evolution of the Weyl parameter, 
which is given by 
\beq
{d\hC\over d\t}=2\alpha a^4\hrho^2\left(1+\hrho-\hH\right).
\label{Cdot}
\eeq
It is obtained by writing $\dot\C=\dot v (d\C/dv)$ and 
by combining (\ref{sigma}), 
(\ref{vdot}) and the junction condition (\ref{fried}). Note that one must 
always have $\hH<1+\hrho$ as a consequence of our previous assumption $f>0$.

The coupled system consisting of (\ref{rhodot}), (\ref{adot}) and (\ref{Cdot})
can be analysed both numerically and, in some specific regimes, analytically.
In the high energy regime $\hrho\gg 1$, the system approaches an asymptotic
behaviour characterized by a rapid growth of the Weyl parameter due to an
abundant production of bulk gravitons: 
\beq
\hC\sim {\alpha\over 4+\alpha} a^4, \quad a\sim t^{1/(4+\alpha)}.
\eeq
One thus sees that the evolution of the scale factor is different 
from the situation  where the graviton outflow is ignored, 
in which case one finds, in the high energy regime,  
 $a\propto t^{1/4}$. However, since $\alpha$ is very small for realistic
cases, the power law is  essentially the same.

In the low energy ($\hrho\ll 1$) radiation era, one recovers  the conventional
evolution $a\sim t^{1/2}$ and the Weyl parameter  approaches a constant value.
Not surprisingly,  this means that, during the late conventional  cosmological
evolution, the production of bulk gravitons can be safely ignored. 

However, what is much more remarkable is that the asymptotic value for $\hC$
is determined quantitatively from the evolution of the brane-bulk system 
in the radiation phase. 
This asymptotic value for $\hC$ can be estimated analytically. If the 
present description is  valid   deep enough in the high energy regime,   
one finds 
\beq\label{rhoweyl}
\epsilon_W\equiv {\rho_{Weyl}\over \rho_{rad}}=
{\hC\over 2 a^4\hrho}\rightarrow {\alpha\over 4},
\eeq
after the high/low energy transition. 
It must be stressed that $\epsilon_W$ depends only on  the number of 
relativistic degrees of freedom at the high/low energy transition, and, 
in particular, 
 {\it is independent of the five-dimensional
Planck mass}.

The ratio  $\epsilon_W$, extrapolated at later times (i.e. 
taking into account the variation of $g_*$), 
is constrained by the number of additional  relativistic 
degrees of freedom allowed during  nucleosynthesis~\cite{osw}, which is 
usually expressed as the number of additional light neutrino species 
$\Delta N_\nu$. A typical bound $\Delta N_\nu\lesssim 1$ implies 
$\epsilon_W\lesssim 8 \times 10^{-2}$ at nucleosynthesis, 
whereas our model yields 
the value
\beq
\epsilon_W \simeq 2\times 10^{-3}\,\left(\frac{\hat{g}\left(T_{\rm
t}\right)}{166.21}\right)\,
\left(\frac{g_*\left(T_{nucl}\right)}{10.75}\right)^{1/3}\,
\left(\frac{106.75}{g_*\left(T_{\rm t}\right)}\right)^{7/3}\,
\eeq
where $T_{\rm t}$ is the temperature at the transition between the high 
and low energy regimes.

It is rather striking that the Weyl contribution which we have estimated 
turns out to be just below the present observational bounds. Given 
that forthcoming observations like the precise measurement of 
CMBR anisotropies might improve significantly the above bound 
\cite{ldht99}, this opens 
the fascinating perspective that the simplest category 
of brane cosmology models could be confronted soon to observational 
tests at the level of {\it homogeneous} cosmology, and not only 
from cosmological perturbations.

\end{document}